# Density Distribution in the Liquid Hg-Sapphire Interface

Meishan Zhao and Stuart A. Rice

The James Franck Institute and Department of Chemistry,

The University of Chicago, Chicago, Illinois, 60637

**Abstract**

We present the results of a computer simulation study of the liquid density distribution normal to the interface between liquid Hg and the reconstructed (0001) face of sapphire. The simulations are based on an extension of the self-consistent quantum Monte Carlo scheme previously used to study the structure of the liquid metal-vapor interface. The calculated density distribution is in very good agreement with that inferred from the recent experimental data of Tamam et al (J. Phys. Chem. Lett. **1**, 1041-1045 (2010)). We conclude that, to account for the difference in structure between the liquid Hg-vapor and liquid-Hg-reconstructed (0001) $Al_2O_3$ interfaces, it is not necessary assume there is charge transfer from the Hg to the $Al_2O_3$. Rather, the available experimental data are adequately reproduced when the van der Waals interactions of the Al and O atoms with Hg atoms and the exclusion of electron density from $Al_2O_3$ via repulsion of the electrons from the closed shells of the ions in the solid are accounted for.

# I. Introduction

Our understanding of the structure of the liquid-vapor interfaces of pure metals and of alloys has advanced considerably in the last few decades, via a combination of theory and simulation [1-19], and grazing incidence x-ray diffraction and x-ray reflectance studies [20-32]. In the most widely applied theoretical description [5-20] the key elements are the construction of a pseudo-potential representation of a multi-component (two-component in the pure metal case), spatially non-uniform, partially disordered distribution of ion cores and valence electrons, and implementation of a self-consistent Monte Carlo simulation procedure using that pseudo-potential representation of the system Hamiltonian to calculate the density distribution and the pair correlation function along the normal and parallel to the liquid-vapor interface of the system. This theoretical analysis generates predictions for the density distribution along the normal to the interface, the in-plane pair distribution function, the magnitude of segregation of solute in the liquid-vapor interfaces of alloys, and the occurrence of two-dimensional crystallization in the segregated layer in the liquid-vapor interface in some dilute alloy systems, e.g. Pb in Ga, all of which in very good agreement with available experimental data.

In contrast with our understanding of the structure of the liquid metal-vapor interface, we have very little understanding of the structures of the liquid-solid interfaces of pure metals and of alloys, since there are few experimental data relevant to the structures of these interfaces and there have been very few theoretical studies that take account of the unique characteristics of the interactions in a liquid metal [33,34]. Specifically, most of these studies have not examined the interplay between the electronic structure of the liquid metal and the solid surface and the effect of that interplay on the density distribution of the liquid side of the interface. To date the overwhelming majority of theoretical studies of the liquid-solid interface have concerned systems with simple short ranged pair interactions, e.g. hard sphere or van der Waals type potentials [35-39]. Both computer simulation and analytical studies of such systems predict that the density distribution in the liquid adjacent to the solid surface is stratified for many layers along the normal to the surface. For the case of a simple fluid in contact with a smooth hard wall the decay length of the amplitude of the density oscillations is predicted to be proportional to the ratio of the amplitude of the first peak of the structure factor to the second derivative of the structure factor evaluated at the first peak [40-46]. When the simple liquid is in contact with a structured wall, e.g. a particular face of a single crystal, the gross features of the interfacial structure remain the same but the penetration depth of the density oscillations and their spacing displays some dependence on the details of the structure of the contacting solid surface [43,44]. Aside from a few studies of the influence of hydrogen bonding between the molecules in the liquid and a solid surface, the possible influence of chemical bonding on the structure of the liquid-solid interfacial region has not been much studied [see ref. 7].

Studies that utilize pair interactions of the types appropriate for simple liquids are not useful for the description of the liquid metal-solid interface in that they fail to capture the possible modification of the interactions between the ion-cores of the liquid metal in the interfacial region that can be induced by the solid surface. It must be expected that the electron density distribution in the liquid metal-solid interface will be altered from that in the liquid metal-vapor interface by virtue of repulsion of electrons from the closed shells of atoms in the solid surface, the response of the electron density distribution to the electrostatic field of the solid surface, exchange interactions with atoms in the solid surface, and charge transfer between the liquid metal and the solid [33,34]. It is also possible that chemical reactions of the atoms of the liquid metal with the surface will occur and thereby both localize some of the electron density and alter the ion-core distribution in the liquid metal side of the interfacial region [7]. The embedded atom potentials used in some simulation studies [45] do take account of the local electron density, but the electron density distribution is assumed to strictly follow the ion-core density distribution and does not allow for the penetration of the electron density outside the liquid metal. In this paper we examine the structure of a liquid metal-solid interface using an extension of the self-consistent quantum Monte Carlo scheme introduced by Rice and coworkers [5-20]. The self-consistency referred to is with respect to the inhomogeneous electron density and ion-core density distributions as described by the Kohn-Sham equation.

We know of very few experimental studies of the density distribution in the liquid metal-solid interface. We consider first experiments designed to study the density distribution along the normal to the liquid metal-solid interface.

Huisman et al [40] have reported a study of x-ray reflectivity from the Ga-diamond (0001) interface. The measured reflectivity as a function of momentum transfer was interpreted with a model that invokes the existence of Ga dimers distributed throughout the liquid, with the density of the layer adjacent to the (0001) surface of diamond 7% greater than the bulk density. Jiang and Rice [7] reinterpreted the measured Ga-diamond reflectivity data. They first showed that a π-complex between diatomic Ga and an ethylenic double bond is stable, and noted that the reconstructed diamond surface has chains of carbon atoms with double bonds. Their self-consistent Monte Carlo simulation results lead to an interface structure with Ga dimers bonded to the reconstructed diamond surface in $\pi$-complexes and Ga atoms in the adjacent liquid. The Ga dimers are present only in the layer bonded to the diamond surface.

Reichert et al [44] have reported a study of x-ray reflectivity from the liquid Pb-Si (001) interface. They interpret their measured reflectivity in terms of an interfacial structure 20 - 30 Å thick and 30% denser than bulk liquid Pb, with the further implication that the enhanced density arises from charge transfer from the Pb to the Si in the amount 0.8 electrons per atom. The x-ray reflectivity from the liquid Pb-$Al_2O_3$ interface does not reveal any enhancement of the density of the liquid Pb in the interfacial region.

In both the Ga-diamond and Pb-Si experiments the range of momentum transfer explored was insufficient to reach any peak associated with a layered density distribution with spacing close to the Ga-Ga or Pb-Pb nearest neighbor separations. The most recent and most extensive study of the liquid metal-solid interface, reported by Tamam et al [43], examines the liquid Hg-$Al_2O_3$ (0001) interface; these experiments do explore momentum transfers large enough to probe the Hg-Hg separation region. Tamam et al interpret their experimental data to show that the layering in the interface is different from that in the liquid Hg-vapor interface. Specifically, they infer that the layering in the liquid Hg-$Al_2O_3$ (0001) interface decays into the liquid more rapidly than is the case for the liquid Hg-vapor interface, and that the amplitude of the first peak in the density distribution is smaller than that of the second peak, which is the reverse of the relative peak amplitudes in the liquid Hg-vapor interface. They also conclude that the average density of the interfacial region is about 10 ± 5% greater than that of bulk Hg. The latter finding they attribute to charge transfer of 0.13 ± 0.05 electron/atom from the Hg to the $Al_2O_3$.

Oh et al [46] have reported a high-resolution transmission electron microscopy study of the liquid Al-$Al_2O_3$ interface. They find the liquid Al in the interface to be stratified with the layer nearest the $Al_2O_3$ less dense than the second layer. As for the liquid Hg-$Al_2O_3$ interface, the amplitude of the first peak in the density distribution is smaller than that of the second peak, which is the reverse of the relative peak amplitudes in the simulated liquid Al-vapor interface.

Consider now experiments designed to study the in plane atomic distribution in the plane of the liquid metal-solid interface. With one exception, these studies deal with monolayers of metals on crystal substrates. It is not at all clear that the monolayer studies provide information that is transferable to interpretation of the structure of the bulk liquid-solid interface. With this caveat in mind, we note that Grey et al [47] have reported a study of x-ray scattering from the interface between a monolayer of liquid Pb and the (111) face of crystal Ge; the data imply that the Pb-Pb spacing is significantly different from that in bulk liquid Pb. Reedijk et al [48, 49] have examined the x-ray scattering from the interface between a monolayer of liquid Sn and the (111) face of Ge. They interpret the diffraction pattern to show the coexistence in the monolayer of liquid-like Sn composed of Sn atoms free to move on the surface and an ordered array of Sn atoms in three-fold sites of the (111) Ge crystal face. The only publication of which we are aware that examines the in-plane structure of a bulk liquid metal-solid interface, by Reichert et al [44], interprets the grazing incidence x-ray diffraction from the Pb-Si interface as indicating the presence of five-fold local order in the liquid layer adjacent to the Si surface.

In this paper we report the results of a study of the density distribution along the normal in the liquid Hg-$Al_2O_3$ (0001) interface using, as already mentioned, an extension of the self-consistent quantum Monte Carlo scheme introduced by Rice and coworkers. Our interest is focused on how and why the structure of the liquid metal-solid interface differs from that of the liquid metal-vapor interface.

## II. Theoretical Background

### A. System Hamiltonian

The pseudo-potential theory representation of the system Hamiltonian is

$$H = \sum_{i=1}^{n} \frac{\mathbf{p}_i^2}{2m} + V_{eff}(R, n_e(\mathbf{r})) + U_0[\rho_0(\mathbf{r}), n_e(\mathbf{r})], \quad (1)$$

where $\mathbf{p}_i$ is the momentum of the $i^{th}$ atom with mass $m$, and $V_{eff}(R, n_e(\mathbf{r}))$ is the collection of effective pair potentials, such that

$$V_{eff}(R, n_e(\mathbf{r})) = \sum_{I=1}^{N} \sum_{J<I}^{N} \phi_{eff}\left(\mathbf{R}_i - \mathbf{R}_j; n_e(\mathbf{r})\right) \quad (2)$$

with $\phi_{eff}\left(\mathbf{R}_i - \mathbf{R}_j; n_e(\mathbf{r})\right)$ the effective pair potential between atoms $i$ and $j$ and

$R = |\mathbf{R}_i - \mathbf{R}_j|$ the distance between atom $i$ and atom $j$. The term $U_0[\rho_0(\mathbf{r}), n_e(\mathbf{r})]$ in Eq.(1) is a structure independent contribution to the energy which is dependent on the electron density $n_e(\mathbf{r})$ and a reference jellium density $\rho_0(\mathbf{r})$. Specifically, the structure independent energy has the form

$$\begin{aligned} U_0[\rho_0(\mathbf{r}), n_e(\mathbf{r})] = & \frac{3(3\pi^2)^{2/3} \sigma}{10} \int_0^\infty [n_e(\mathbf{r})]^{5/3} dz + \frac{\sigma}{72} \int_0^\infty \frac{|\nabla^2 n_e(\mathbf{r})|}{n_e(\mathbf{r})} dz \\ & + \frac{\sigma}{540(3\pi^2)^{2/3}} \int_0^\infty dz [n_e(\mathbf{r})]^{1/3} \left[ \left(\frac{\nabla^2 n_e(\mathbf{r})}{n_e(\mathbf{r})}\right)^2 \right. \\ & \left. - \frac{9}{8}\left(\frac{\nabla^2 n_e(\mathbf{r})}{n_e(\mathbf{r})}\right)\left(\frac{\nabla n_e(\mathbf{r})}{n_e(\mathbf{r})}\right)^2 + \frac{1}{3}\left(\frac{\nabla n_e(\mathbf{r})}{n_e(\mathbf{r})}\right)^4 \right] \\ & - 2\pi\sigma \int_0^\infty dz \int_0^\infty dz' [\rho_0(z)\rho_0(z') - n_e(z)n_e(z')]|z-z'| \\ & + 2\sigma \int_0^\infty n_e(z)\varepsilon_{xc}[n_e(z)]dz \\ & + 2\sigma \int_0^\infty \rho_0(z)\varepsilon_{ps}[n_e(z)]dz. \end{aligned} \quad (3)$$

Detailed discussions of the various terms in Eq. (3) can be found in our earlier work and the research literature [8-19, 49-55].

The last term in the equation above is the so called electron-ion pseudo-potential

$$\varepsilon_{ps} = \frac{3}{k_F^3}\int_0^{k_F} f(q,q)q^2 dq - \frac{1}{\pi}\sum_{i=1}^{2} X_i \int_0^{\infty} dq \left[ z_i^{*2}|M_i(q)|^2 - (z_i^*)^2 F_{ii}(q) \right]$$
$$- \frac{2\pi}{\Omega}\sum_{i=1}^{2}\sum_{j=1}^{2} z_i^* z_j^* X_i X_j \lim_{q \to 0}\frac{1-F_{ij}(q)}{q^2} + \left(\frac{\alpha}{r_s} + \frac{\beta}{r_s}\right) \quad (4)$$

where $k_F$ is the Fermi wave number, $f(q, q)$ is the diagonal matrix element of the Fourier transform of the nonlocal bare electron-ion pseudo-potential $V_{ps}^{ion}(r)$, $M_i(q)$ is the Fourier transform of the depletion hole distribution, $X_i$ is the mole fraction of the $i^{th}$ component and $r_s$ is the radius of a sphere per unit electron that is determined from $4\pi r_s^3/3 = (n_{e,bulk})^{-1}$, where $n_{e,bulk}$ is the bulk electron density of the liquid.

**B. Electron-ion pseudo-potential**

For the calculations reported in this paper we adopted the non-local energy independent model pseudo-potential proposed by Woo, Wang and Matsuura [56] This potential has been previously used by Rice and coworkers in studies of the structures of the liquid-vapor interfaces of various pure liquid metals and binary alloys. It has the form

$$V_{ps}^{ion}(r) = \sum_l \left[\overline{V}_l(r) + (V_{1l}(r) - \overline{V}_l(r))R_{1l}\rangle\langle R_{1l}|\right]|l\rangle\langle l|, \quad (5)$$

where $\overline{V}_l(r)$ is a pseudo-potential average over all states other than the first valence state for a given angular momentum $l$; $\overline{V}_l(r)$ is calculated in the same way as is $V_{1l}(r)$ except with replacement of the parameter $B_{1l}$ by $\overline{B}_{1l}$. $|R_{1l}\rangle$ is the radial component of the wave function for the state $|1l\rangle$, and $|l\rangle$ is a simple projection onto the state with angular momentum $l$. With these definitions the model pseudo-potential takes the form

$$V_{1l}(r) = \begin{cases} -B_{1l} + Z_l/r, & r \leq R_l \\ Z_l/r, & r > R_l \end{cases} \quad (6)$$

where $B_{1l}$, $Z_l$ and $R_l$ are parameters; they are usually determined by a pseudo-eigenfunction expansion and perturbation theory. $Z$ is the valence of the ion. The state $|1l\rangle$ can be separated into radial and angular parts, such that $\langle x|1l\rangle = (A/r)y_{1l}(r)Y_{lm}(\theta,\phi)$, where A is a normalization constant, $Y_{lm}(\theta,\phi)$ is a spherical harmonic function, and $y_{1l}(r)$ is the radial wave function given by

$$y_{1l}(r) = \begin{cases} M_{\nu,l+1/2}(2\lambda r), & r \leq R_l \\ W_{\nu_0,l+1/2}(2\lambda_0 r), & r > R_l \end{cases} \quad (7)$$

where $M_{\nu,l+1/2}(2\lambda r)$ and $W_{\nu_0,l+1/2}(2\lambda_0 r)$ are, respectively, the regular and irregular Whittacker functions. The parameters of the Whittacker functions are $\lambda_0 = \sqrt{-2E_{1l}}$, $\lambda = \sqrt{-2(E_{1l} + B_{1l})}$, $\nu_0 = Z_l/\lambda_0$, $\nu = -Z_l/\lambda$, with $E_{1l}$ the spectroscopic term energy of the state $|1l\rangle$. The potential parameters we have used for the calculations reported in this paper are given in Table 1.

**C. Ion-ion pair potential**

The effective ion-ion pair potential is calculated using a local density approximation

$$\phi_{eff}(R, n_e(\mathbf{r})) = \phi_H\left(R; \frac{1}{2}[n_e(\mathbf{r}_i) + n_e(\mathbf{r}_j)]\right). \tag{8}$$

In a homogeneous liquid metal with electronic density $n_e(\mathbf{r})$ the pair interaction is calculated from

$$\phi_H(R) = \frac{z_i^* z_j^*}{R}\left[1 - \frac{1}{\pi}\int_0^\infty [F_{ij}(q) + F_{ji}(q)]\frac{\sin(qR)}{q}dq\right] + \phi_{BM}(R) + \phi_{vw}(R). \tag{9}$$

That interaction includes a direct Coulomb repulsion between ions with effective valence charges $z_i^*$ and $z_j^*$; an indirect interaction mediated by the conduction electrons, i.e. the band structure energy; the Born-Mayer core-core repulsion [50], $\phi_{BM}(R)$; and the van der Waals polarization interaction between ion cores, $\phi_{vw}(R)$. Specifically, $\phi_{BM}(R) = Ae^{-BR}$, where A and B are the Born-Mayer potential parameters. Both $\phi_{BM}(R)$ and $\phi_{vw}(R)$ are generally much smaller than the other terms contributing to the energy of the liquid metal.

The normalized energy wave-number characteristic function is, as derived by Shaw [51],

$$F_{ij}(q) = \frac{\Omega^2 q^4}{16\pi^2 z_i^* z_j^*}\left\{\frac{[1-\varepsilon(q)](v_1+v_2)^2}{\varepsilon(q)} + 2g(q)(v_1+v_2) + \varepsilon(q)g(q)^2 + h(q)\right\} \tag{10}$$

where $\Omega = \rho_{bulk}^{-1}$ is the volume per atom, $\rho_{bulk}$ is the bulk liquid density, $\varepsilon(q)$ is the wave-number dependent Hartree dielectric function. The local potential contributions $v_1$ and $v_2$, arising from the valence charge $Z$ and the depletion hole charge $\bar{Z}$, are given by

$$v_1 = -\frac{2\pi}{\Omega q^2}(Z_i + Z_j), \tag{11}$$

$$v_2 = -\frac{2\pi}{\Omega q^2}(\bar{Z}_i + \bar{Z}_j). \tag{12}$$

The nonlocal screening function *g(q)* and the nonlocal bare pseudopotential contribution *h(q)* to the second order approximation are given by

$$g(q) = \frac{4}{\pi^2 q^2 \varepsilon(q)} \int_{k \leq k_F} \frac{f(\mathbf{k},\mathbf{q})}{k^2 - |\mathbf{k}+\mathbf{q}|^2} d\mathbf{k}, \tag{13}$$

$$h(q) = \frac{4}{\pi^2 q^2} \int_{k \leq k_F} \frac{|f(\mathbf{k},\mathbf{q})|}{k^2 - |\mathbf{k}+\mathbf{q}|^2} d\mathbf{k}, \tag{14}$$

and *f(k,q)* is the matrix element of the Fourier transform of the nonlocal bare electron-ion pseudo-potential $V_{ps}^{ion}(r)$.

The normalized energy wave-number characteristic functions are shown in Fig. 1 for liquid Hg for a few selected densities. The various terms contributing to the surface energies are shown in Fig. 2. The effective ion pair interaction is displayed in Fig. 3 at the experimental liquid Hg density, 0.0407 at./Å$^3$.

### D. Character of the jellium background

One of the important issues that must be addressed in the pseudo-potential theory is an appropriate selection of the jellium background. It defines a reference atomic density profile in the pseudo-atom Hamiltonian. As in our previous work we chose the functional form [5-6,10-19]

$$\rho(z, z_0, \beta) = \rho_{bulk} \left[ 1 + \exp\left(\frac{|z| - z_0}{\beta}\right) \right]^{-1}. \tag{15}$$

Eq. (15) represents the charge distribution of a slab of ions with two surfaces perpendicular to the z axis, where $z_0$ is the position of the Gibbs dividing surface and $\beta$ measures the width of the inhomogeneous region of the profile. This distribution is normalized by setting $\rho_{bulk} = N/2\sigma z_0$, where $N$ is the total number of atoms in the slab and $\sigma$ is the area of the slab. The parameters $z_0$ and $\beta$ are varied to obtain the best fit to the instantaneous ionic configuration. The electron density obtained by using the effective valence representation may deviate to some extent from the true electron density. The needed correction can be obtained, using the simulated atomic density as the input, by iteratively re-solving the Kohn-Sham equation [57, 58] self-consistently. In the current work, the electron density distribution is obtained from self-consistent solutions to the Kohn-Sham equation as proposed by Eguiluz et al [59-61],

$$\left[-\frac{h^2}{8\pi^2 m}\nabla^2 + V_{eff}(\mathbf{r})\right]\psi_n(\mathbf{r}) = E_n \psi_n(\mathbf{r}), \tag{16}$$

where $V_{eff}(\mathbf{r})$ is an effective potential,

$$V_{eff}(\mathbf{r}) = V_{ps}(\mathbf{r}) + V_{xc}(\mathbf{r}) + \int d\vec{r} \frac{n_e(\mathbf{r})}{|\mathbf{r}-\mathbf{r}'|} \tag{17}$$

including the electron-jellium background pseudo-potential interaction $V_{ps}(\mathbf{r})$, the exchange-correlation potential $V_{xc}(\mathbf{r})$, and the electrostatic potential. The electron number density is then obtained from

$$n_e(\mathbf{r}) = \sum_{n=1}^{\infty} f_n |\psi_n(\mathbf{r})|^2, \tag{18}$$

with $f_n$ the electron occupation number in the quantum state $\psi_n(\mathbf{r})$.

In our slab geometry the ion density distributions in the $x$ and $y$ directions are uniform, hence the electronic wave function $\psi_n(\mathbf{r})$ takes the form

$$\psi_n(\mathbf{r}) = e^{i(k_x x + k_y y)} \phi_n(z), \tag{19}$$

where $k_x$ and $k_y$ are the wave numbers in the $x$ and $y$ directions, respectively. The electron density is then a function only of position along the normal to the interface,

$$n_e(\vec{r}) = \sum_{v=1}^{\infty} f_v |\phi_n(z)|^2. \tag{20}$$

With the results of this calculation we construct the electron density dependent potential $V_{eff}(z; n_e(z))$, and iterate the procedure until a self-consistent solution is obtained. We show in Fig. (4) the typical mean valence electron density imposed on a jellium reference density for liquid Hg, calculated as described above.

**E. Al$_2$O$_3$ surface**

The 0001 surface of Al$_2$O$_3$ is reconstructed relative to the atomic positions generated when the bulk crystal is cleaved to expose that surface [62]. A top view and a side view of that surface are displayed in Fig. 5. Noting that the Al and O atoms in the reconstructed surface are almost coplanar, and that the surface is stoichiometric, we argue that charge and dipolar interactions between the liquid Hg and Al$_2$O$_3$ terminated in this plane can be neglected. And, since Al$_2$O$_3$ has a very large band gap we expect that electron transfer from Hg to the solid can be neglected. We then represent the interaction between 0001 terminated Al$_2$O$_3$ and liquid Hg as the sum of van der Waals interactions between Hg, Al and O and a soft repulsion to account for the exclusion of the electron density distribution from the closed shells of the Al and O ions [63-66],

$$U(R) = -\frac{3}{2}\left(\frac{I_1 + I_2}{I_1 I_2}\right)\frac{\alpha_1 \alpha_2}{(4\pi\varepsilon_0)^2}\frac{1}{R^6} + A_{12}e^{-b_{12}R}$$

with $A_{12} = (A_1 A_2)^{1/2}$, $b_{12} = (b_1 + b_2)/2$. And $I_j$ is the first ionization energy of atom $j$, $\alpha_j$ is the polarizability of atom $j$, $\varepsilon_0 = 8.854 \cdot 10^{-12} C^2 J^{-1} m^{-1}$ is permittivity of free space, and $A_j$ and $b_j$ are the Born-Mayer potential parameters for atom $j$. These potential parameters are listed in Table 2. The potential is plotted in Fig.6.

### III. Simulation methodology

Our simulations were carried out on a Dell-Linux computer cluster. Since the Hg ions are mobile, we need to calculate a suitable average over all allowed ionic configurations in order to determine the longitudinal and transverse density distributions in liquid Hg and in the liquid Hg-sapphire interface. As discussed in the last section, an initial jellium distribution is used to generate an electronic density distribution from which the ion-electron pseudo-potential and effective ion-ion interaction potentials are calculated. The initial jellium distribution is then used in a Monte Carlo simulation of the inhomogeneous liquid Hg-sapphire interface. Since each Monte Carlo step changes the ion distribution, it also changes the electronic density distribution, hence the ion-electron pseudo-potential and the effective ion-ion interaction; this effect is particularly important in the liquid-sapphire interface region. Accordingly, when the ion distribution is changed, the electron distribution is recalculated, to be consistent with the new ion distribution; this procedure is continued until the Monte Carlo simulation converges.

The simulation system consisted of a slab of 1000 Hg atoms, with dimensions of $L_0 \times L_0 \times 2L_0$ in the *(x, y, z)* directions, in contact with the reconstructed $Al_2O_3$ (0001) surfaces, as shown in Fig. 5, in the positive *z* and negative *z* directions, so that the area of each of the two Hg-sapphire interfaces is $\sigma = L_0^2$. The size of the slab, $L_0$, was chosen such that the average density of ions in the slab matched the density of liquid Hg at the simulation temperature. The slab holds about 8 layers of Hg with, on average, about 120 Hg atoms in contact with the sapphire wall.

In our simulations, the center of mass of the system was located at the origin of the coordinates (*x = 0, y = 0, z = 0*). Periodic boundary conditions are applied in the *x* and *y* directions, and in the ±*z* directions far away from the Hg/sapphire interfaces. The initial ion configuration was generated by placing the particles within the boundaries of the slab, subject to the constraint that no ion-ion separation was less than the ionic diameter.

As in our earlier reported calculations, we have adopted an efficient computational strategy and data management scheme for the calculations. Prior to starting the simulation we compute and tabulate the effective ion-ion interaction potential energies for a series of electron densities ranging from somewhat below to somewhat above the bulk density of liquid Hg. During the simulation the interaction between a particular pair of ions is obtained from a rational functional interpolation for the given electron density using the pre-calculated data bank. The simulations were carried out using the Metropolis scheme [67] and a force bias Monte Carlo algorithm to eliminate the overlaps between ion cores. The trial configurations were generated by randomly displacing a selected ion-core; the magnitude of the ion-core displacement was chosen to lead to convergence to equilibrium with a reasonable overall acceptance ratio for the trial configurations. Each iteration (each pass) in our simulation has 1000 moves, with one move per Hg ion-core. To ensure sampling the equilibrium configurations of the ion-cores, we eliminate several tens of thousands of passes from the initial stages of the simulation when collecting the data for final statistics.

## IV. Results

**A. The transverse pair correlation function**

As in past work we characterize the structure of the liquid-solid interface with the transverse (in-plane) pair correlation function and the longitudinal density distribution. In our calculation the transverse pair correlations function is calculated from a histogram of the separations of all pair of particles in a thin slice of the interfacial region in the normalized correlation function form

$$g(r) = \frac{2V_T N(r, \Delta r)}{V_s N_T^2}$$

where $N_T$ is the total number of particles in the slice, $N(r, \Delta r)$ is the average number of pairs of particles between r and $(r + \Delta r)$ and in the slice, $V_T$ is the total volume of all the particles in the slice, and $V_s$ is the average volume of the intersection of the spherical shell between r and $(r + \Delta r)$ of the thin slice.

The pair correlation function of bulk liquid Hg has been measured many times. To calibrate the accuracy of our simulations we show in Fig.7 a comparison of the simulated pair correlation function with the experimentally determined pair correlation function at $T = 293$K with density 0.0407 at./angstrom$^3$ [68,69]. The agreement between the simulation data and the experimental data is very good.

**B. The longitudinal density distribution**

Comparison of the predicted and experimentally inferred longitudinal density distributions in the liquid Hg/sapphire interface provides the most sensitive test of our calculations. The major aspects of the longitudinal density distribution reported by Tamam et al [43] are: (1) The liquid Hg-$Al_2O_3$ interface is stratified. This result differentiates this interfacial system from the Ga-diamond and Pb-Si interfacial systems. (2) The layer of Hg in the interface closest to the $Al_2O_3$ surface has smaller amplitude than the amplitude of the outermost peak in the stratified liquid Hg-vapor interface. Moreover, the amplitude of the second layer is much greater than that of the first layer, whereas the reverse is the case for the Hg-vapor interface. (3) The average density of the interfacial region is about $10 \pm 5\%$ greater than that of bulk Hg. Using an empirical relation between ionization state and atomic radius Tamam et al [43] infer that there is a charge transfer of $0.13 \pm 0.05$ electron/atom from the Hg to the $Al_2O_3$.

We have calculated the longitudinal density distribution of the ions in the liquid Hg-$Al_2O_3$ interface from a histogram of the distance between a particle and the center of mass of the slab; the density profiles were averaged from the opposite halves of the slab to obtain the reported density distribution. Fig. 8 shows a normalized longitudinal density profile from calculation after 92000 passes. A comparison to experiment observation along aside the Hg/vapor interface density profiles is given in Fig.9 which shows that the calculated longitudinal density profile agrees well with the experimental density profile.

As in our previous study of the longitudinal density distribution in the liquid-vapor interface of Hg, the temperature dependences of the amplitudes and widths of the peaks in the longitudinal density distribution are not very sensitive to temperature. Specifically, there is no noticeable difference between the simulation results for 293K and 285K.

## IV. Discussion

The self-consistent quantum Monte Carlo simulations of the longitudinal density distributions in the liquid Hg-vapor interface and in the liquid-Hg-reconstructed (0001) $Al_2O_3$ interface reported in this paper are both in very good agreement with the available experimental data. The simulated longitudinal density distributions reproduce both the general similarities and the general differences between these distributions in the two interfaces. In particular, the layer of the stratified liquid-Hg-reconstructed (0001) $Al_2O_3$ interface adjacent to $Al_2O_3$ is less dense than is the corresponding layer in the liquid Hg-vapor interface, and second layer of the stratified liquid-Hg-reconstructed (0001) $Al_2O_3$ interface is denser than the first layer, which ratio is opposite that found for the liquid Hg-vapor interface. The excess density in the stratified liquid-Hg-reconstructed (0001) $Al_2O_3$ interface is well accounted for. All of these results follow from a common treatment of the electronic structure of the liquid metal. The difference between the longitudinal density distributions in the liquid Hg-vapor and liquid-Hg-reconstructed (0001) $Al_2O_3$ interfaces arises entirely from the van der Waals interactions of the Al and O atoms with Hg atoms and the exclusion of electron density from $Al_2O_3$ via repulsion of the electrons from the closed shells of the ions in the solid. We conclude that, to account for the difference in structure between the liquid Hg-vapor and liquid-Hg-reconstructed (0001) $Al_2O_3$ interfaces it is not necessary assume there is charge transfer of $0.13 \pm 0.05$ electron/atom from the Hg to the $Al_2O_3$. Rather, the available experimental data are adequately reproduced when the van der Waals interactions of the Al and O atoms with Hg atoms and the exclusion of electron density from $Al_2O_3$ via repulsion of the electrons from the closed shells of the ions in the solid are accounted for.

## V. Acknowledgments

The research reported in this paper has been supported by a grant from the Department of Energy ER46321.

**Figure Captions:**

Fig. 1. The normalized energy wave-number characteristic functions at selected densities of liquid Hg, 0.01045, 0.04070, and 0.06792 at./Å$^3$ respectively.

Fig. 2. Contributions to the surface energy as a function of the longitudinal positive ion jellium density parameter: (a) electrostatic energy, exchange-correlation energy, and pseudo-potential energy; (b) kinetic energy and total surface energy.

Fig. 3. The effective ion-ion pair interaction potential in liquid Hg at the experimental bulk density 0.0407 at./Å$^3$.

Fig. 4. Mean valence electron density profile in liquid Hg superposed on the jellium reference density at liquid density 0.0407 at./Å$^3$: (a) $\beta = 0.1$ and (b) $\beta = 2.0$.

Fig. 5. A reconstructed (0001) Al$_2$O$_3$ surface with relative atomic positions generated when the bulk crystal is cleaved to expose that surface. The upper panel is a top view and the lower panel is a side view [from Ref. 62]. It is noted that the Al and O atoms in the reconstructed surface are almost coplanar with $d_{\perp 1} = 0.1$Å.

Fig. 6. The effective potentials between Hg, Al and O atoms at the Hg/sapphire interface. These potentials account for both the van der Waals interactions and the exclusion of the electron density distribution from the closed shells of the Al and O ions.

Fig. 7. Transverse pair correlation function of bulk liquid Hg from the self consistent Monte Carlo simulations at a density of 0.04079 at./Å$^3$ and $T = 293$K, with comparison to the experimental observations reported by Waseda et al [68] and by Boiso et al [69].

Fig. 8. Normalized longitudinal density profiles in the liquid Hg-sapphire interface from the self-consistent Monte Carlo simulations.

Fig. 9. Comparison of the simulated and experimental longitudinal density profiles in the liquid Hg-reconstructed 0001 $Al_2O_3$ interface, and in the liquid Hg-vapor interface.

Table 1. Ionic pseudo-potential parameters (in units of a.u.) for Hg. $r_{max}$ is the maximum value of $r$ in the radial wave function

| $l$ | $E_{1l}$ | $R_l$ | $B_{1l}$ | $\bar{B}_{1l}$ | $r_{max}$ |
|---|---|---|---|---|---|
| 0 | 0.6890786 | 1.8 | 1.180163 | 0.700763 | 35.0 |
| 1 | 0.4545650 | 1.7 | 1.119794 | 0.426535 | 45.0 |
| 2 | 0.2108822 | 3.0 | 0.669270 | 0.638592 | 55.0 |

Table 2. Potential parameters of Hg-sapphire interaction (in units of a.u.): $A$ and $b$ are the Born-Mayer potential parameters, $I$ is the first ionization energy, and $\alpha(a_0^3)$ is the polarizability.

|  | $Hg^{2+}$ | $Al^{3+}$ | $O^{2-}$ | Hg | Al | O |
|---|---|---|---|---|---|---|
| $\alpha$ | 10.53 | 5.23 | 13.56 |  |  |  |
| $I$ |  |  |  | 1.602 | 0.918 | 2.089 |
| $A$ |  |  |  | 2230.0 | 157.85 | 78.771 |
| $b$ |  |  |  | 1.85254 | 1.94681 | 2.00474 |

**References:**

1. S. A. Rice, J. Non-Cryst. Solids **207**, 755 (1996).

2. M. P. D'Evelyn and S. A. Rice, Phys. Rev. Lett. **47**, 1844 (1981).

3. M. P. D'Evelyn and S. A. Rice, J. Chem. Phys. **78**, 5081 (1983).

4. M. P. D'Evelyn and S. A. Rice, J. Chem. Phys. **78**, 5225 (1983).

5. J. G. Harris, J. Gryko and S. A. Rice, J. Chem. Phys. **87**, 3069 (1987).

6. J. G. Harris, J. Gryko and S. A. Rice, J. Stat. Phys. **48**, 1109 (1987).

7. X. Jiang and S. A. Rice, J. Chem. Phys. **123**, 104703 (2005).

8. X. Jiang, M. Zhao and S. A. Rice, Phys. Rev. B **72**, 094201 (2005).

9. X. Jiang, M. Zhao and S. A. Rice, Phys. Rev. B **71**, 1042031 (2005).

10. M. Zhao and S. A. Rice Phys. Rev. B. **63**, 85409 (2001).

11. S. A. Rice and M. Zhao, J. Phys. Chem. A **103**, 10159 (1999).

12. M. Zhao, D. Chekmarev and S. A. Rice, J. Chem. Phys. **108**, 5055 (1998).

13. S. A. Rice and M. Zhao, Phys. Rev. B **57**, 13501 (1998).

14. M. Zhao, D. Chekmarev, Z. Cai and S. A. Rice, Phys. Rev. E **56**, 7033 (1997).

15. S. A. Rice, M. Zhao and D. Chekmarev, in *Microscopic Simulation of Interfacial Phenomena in Solids and Liquids,* edited by S.R. Phillpot, P.D. Bristowe, D.G. Stroud, J.R. Smith, (The Materials Research Society, Vol. 492, 1998), pp.3-14.

16. M. Zhao, D. Chekmarev and S. A. Rice, J. Chem. Phys. **109**, 768 (1998).

17. D. Chekmarev, M. Zhao and S. A. Rice, J. Chem. Phys. **109**, 1959 (1998).

18. D. Chekmarev, M. Zhao and S. A. Rice, Phys. Rev. E **59**, 479 (1999).

19. M. Zhao, D. Chekmarev, Z.-H. Cai and S. A. Rice, Phys. Rev. E **56**, 7033 (1997).

20. D. Li and S. A. Rice, Phys. Rev. E **72**, 41506 (2005).

21. D. Li, X. Jiang X, B. Yang and S. A. Rice, J. Chem. Phys. **122**, 224702 (2005).

22. B. Yang, D. Li and S. A. Rice, Phys. Rev. B **67**, 212103 (2003).

23. B. Yang, D. Li and S. A. Rice, Phys. Rev. B **67**, 54203 (2003).

24. D. Li, B. Yang and S. A. Rice, Phys. Rev. B **65**, 224202 (2002).

25. B. Yang, D. Li, Z. Huang and S. A. Rice, Phys. Rev. B **62**, 13111 (2000).

26. N. Lei, Z. Huang and S. A. Rice, J. Chem. Phys. **107**, 4051 (1997).

27. N. Lei, Z. Huang and S. A. Rice, J. Chem. Phys. **104**, 4802 (1996).

28. N. Lei, Z. Huang and S. A. Rice and C. Grayce, J. Chem. Phys. **105**, 9615 (1996).

29. E. B. Flom, M. Li, A. Acero, N. Marskil and S. A. Rice, Science **260**, 332 (1993).

30. M. J. Regan, P. S. Pershan, O. M. Magnussen, B. M. Ocko, M. Deutsch and L. E. Berman, Phys. Rev. B **55**, 15874 (1997).

31. M. J. Regan, E. H. Kawamoto, S. Lee, P. S. Pershan, N. Maskil, M. Deutsch, O. M. Magnussen, B. M. Ocko and L. E. Berman, Phys. Rev. Lett. **75**, 2498 (1995).

32. O. M. Magnussen, B. M. Ocko, M. J. Regan, K. Penanen, P. S. Pershan and M. Deutsch, Phys. Rev. Lett. **74**, 4444 (1995).

33. Z-h. Cai, J. Harris and S. A. Rice, J. Chem. Phys. **89**, 2427 (1988).

34. Z-h. Cai and S. A. Rice, J. Chem. Phys. **90**, 6716 (1989).

35. W. E. McMullen and D. W. Oxtoby, J. Chem. Phys. **88,** 4146 (1987).

36. W. A. Curtin, Phys. Rev. Lett. **59,** 1228 (1987).

37. J. H. Sikkenk, J. O. Indekeu, J. M. J. van Leeuwen and E. O. Vossnack, Phys. Rev. Lett. **59,** 98 (1987).

38. W.–L. Ma, J. R. Banavar and J. Koplik, J. Chem. Phys. **97,** 485 (1992).

39. M. F. Toney, *et al. Nature* **368,** 444 (1994).

40. W. J. Huisman, J. F. Peters, M. J. Zwanenburg, S. A. de Vries, T. E. Derry, D. Abernathy and J. F. van der Veen, Nature, **390**, 379 (1997).

41. W. J. Huisman, J. F. Peters, J. W. Derks, et al., Rev. Sci. Instr. **68**,4169 (1997).

42. W. J. Huisman, J. F. Peters, S. A. deVries, et al., Surf. Sci. **387**, 342 (1997).

43. L. Tamam, D. Pontoni, T. Hofmann, B. M. Ocko, H. Reichert and M. Deutsch, J. Phys. Chem. Lett. **1**, 1041 (2010).

44. H. Reichert, O. Klein, H. Dosch, M. Denk, V. Honkimaki, T. Lippermann and G. Reiter, Nature **408**, 839 (2000).

45. E. T. Chen, R. N. Barnett and U. Landman, Phys. Rev. **B 40**, 924 (1989).

46. H. Oh, Y. Kauffmann, C. Sheu, W. D. Kaplan and M. Ruhle, Science **310**, 661 (2005).

47. F. Grey, R. Feidenhans, J. S. Pedersen, M. Neilsen and R. L. Johnson, Phys. Rev. B **41**, 9519 (1990).

48. M. F. Reedijk, J. Arsic, D. Kaminski, P. Poodt, H. Knops, P. Serrano, G. R. Gastro and D. Vlieg, Phys. Rev. B **67**, 165423 (2003).

49. M. F. Reedijk, J. Arsic, F. K. Theije, M. T. McBride, K. F. Peters and E. Vlieg, Phys. Rev. B **64**, 33403 (2001).

50. T. L. Gilbert, J. Chem. Phys. **49**, 2640 (1968).

51. R.W. Shaw, Jr., Phys. Rev. **174**, 764 (1968).

52. N. H. March, Adv. Phys. **6**, 1 (1957).

53. D. A. Kirzhnits, Sov. Phys. JETP **5**, 64 (1957).

54. S. H. Vosko, L. Wilk and M. Nusair, Can. J. Phys. **58**, 1200 (1980).

55. D. C. Langreth and M. J. Mehl, Phys. Rev. B **28**, 1809 (1983).

56. C. H. Woo, S. Wang and M. Matsuura, J. Phys. F Metal Phys. **5**, 1836 (1975).

57. W. Kohn and L. J. Sham Phys. Rev. A **140**, 1133 (1965).

58. P. Hohenberg and W. Kohn, Phys. Rev. B **136**, 864 (1964).

59. A. K. Karmakar and R. N. Joarder, Phys. B. **245**, 81 (1998).

60. A. G. Eguiliz, D. A. Campbell, A. A. Maradudin and R. F. Wallis, Phys. Rev. B 30, 5449 (1984).

61. J. P. Perdew, M. Ernerhof, A. Zupan and K. Burke, J. Chem. Phys. 108, 1522 (1998).

62. T. J. Godin and J. P. LaFemina, Phys. Rev. B 49, 7691 (1993).

63. R. S. Berry, S. A. Rice, and J. Ross, *Physical Chemistry*, 2$^{nd}$ Ed., Oxford University Press, 2000. pp. 303, eq.(10.19); D. A. McQuarrie and J. D. Simon, *Physical Chemistry: A Molecular Approach*, University Science Books Publishing, 1 edition, July, 1997. pp.669.

64. N. W. Grimesy and R. W. Grimesz, J. Phys.: Condens. Matter **10**, 3029 (1998).

65. A. A. Abrahamson, Phys. Rev. **178**, 76 (1969).

66. G. Bobel, A. Longinotti, and F. G. Fumi, J. Physique **48**, 45 (1987).

67. N. Metropolis, A. W. Rosenbluth, M. N. Rosenbluth, A. M. Teller and E. Teller, J. Chem. Phys. **21**, 1087 (1953).

68. Y. Waseda, The structure of Non-Crystalline Materials - Liquids and Amorphous Solids, McBraw-Hill, New York, 1980.

69. 33L. Boiso, R. Cortes, and C. Segaud, J. Chem. Phys. **71**, 3595 (1979).

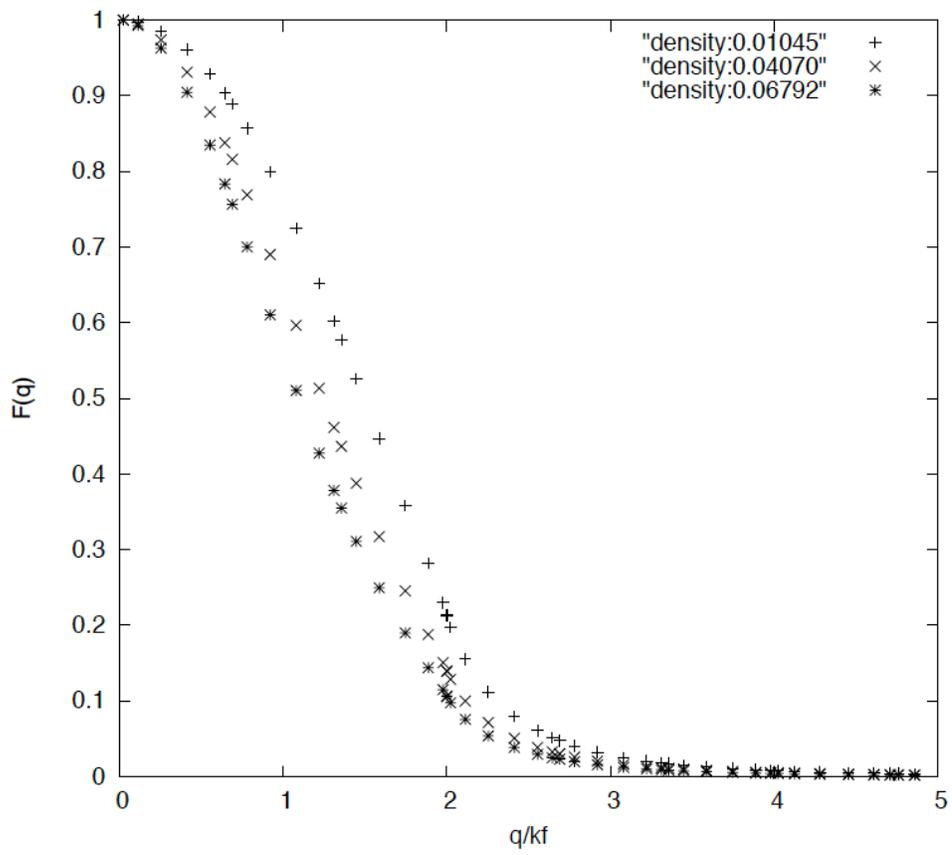

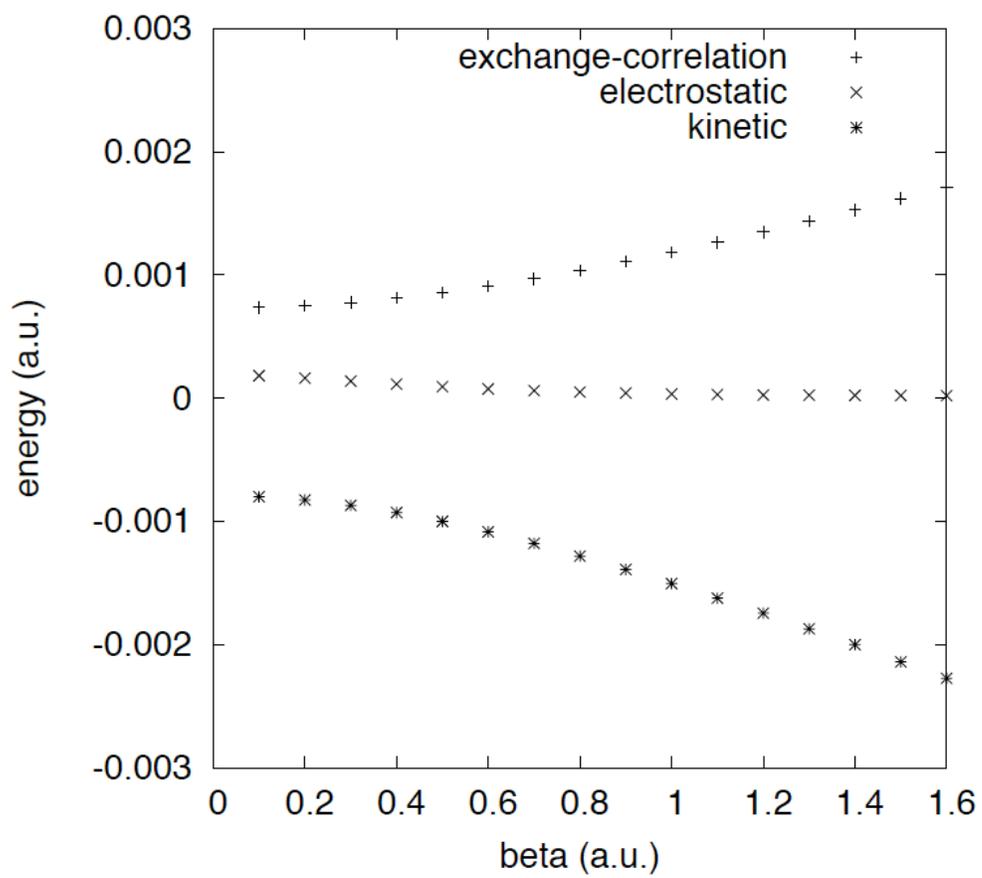

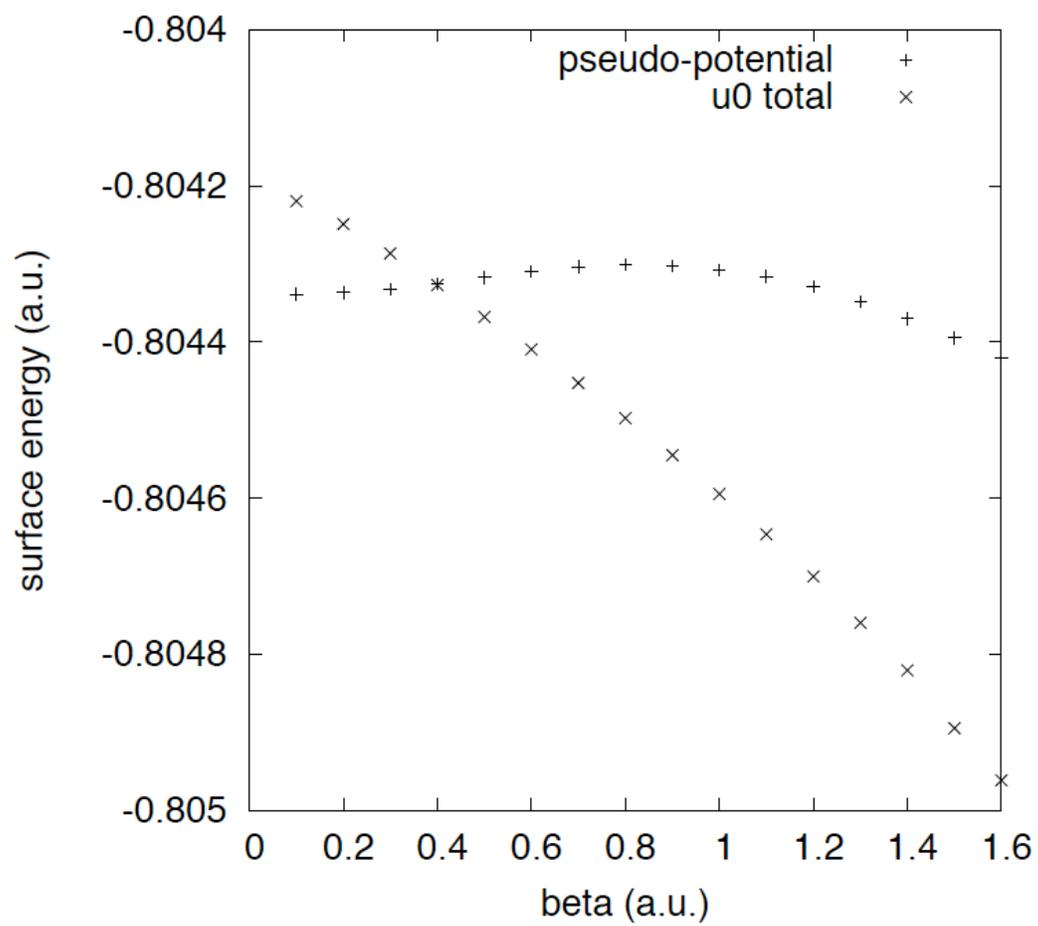

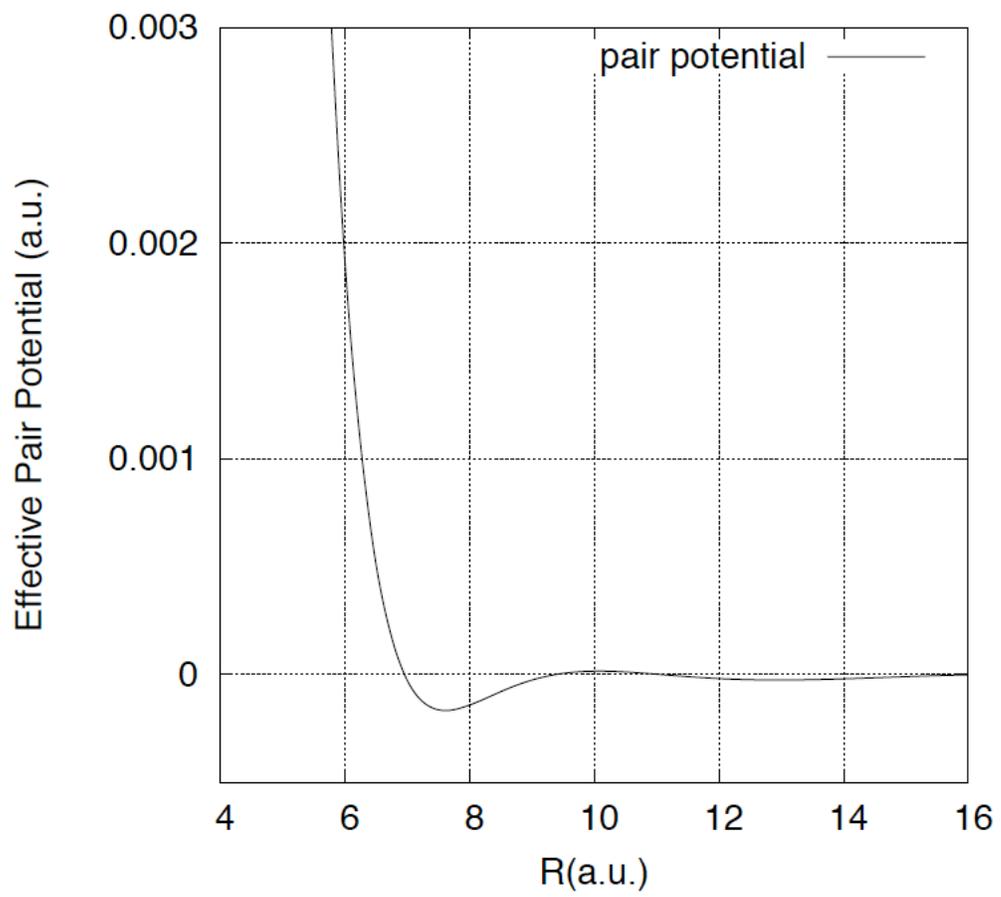

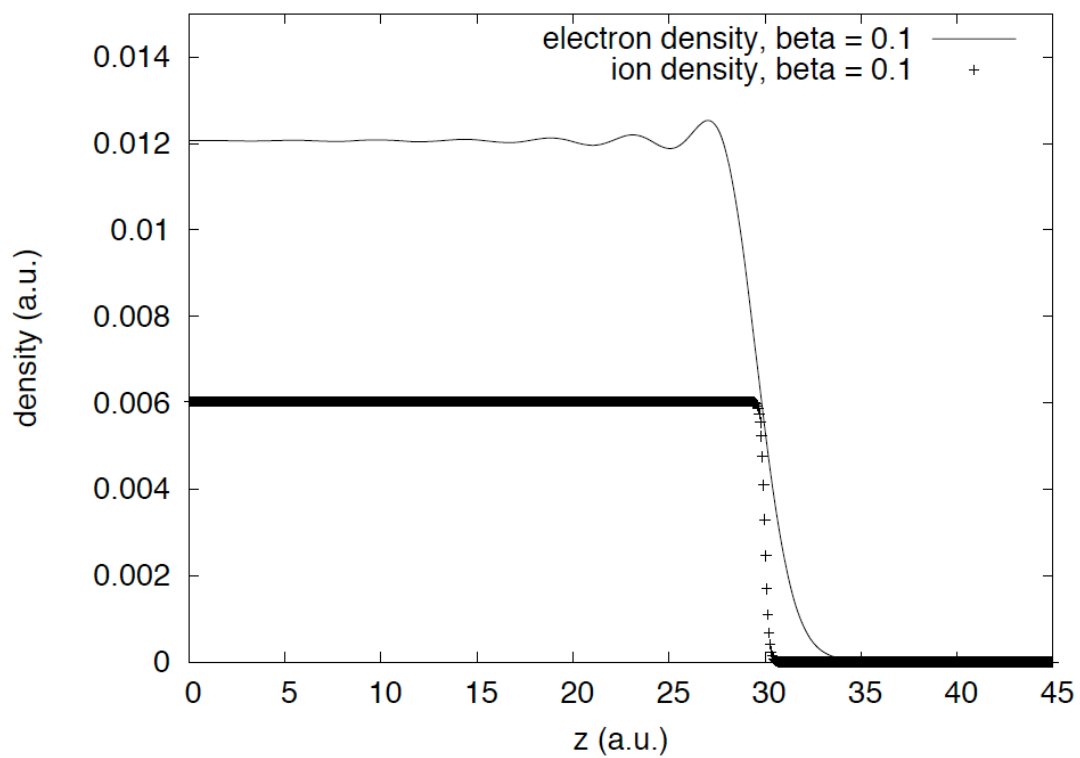

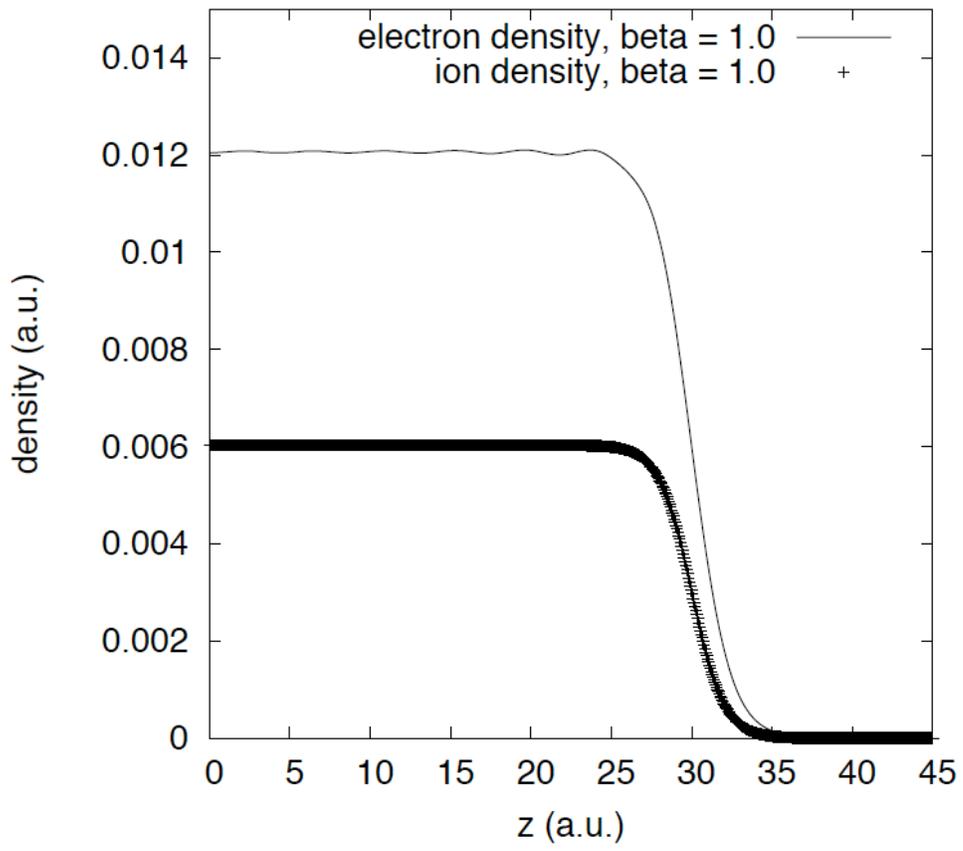

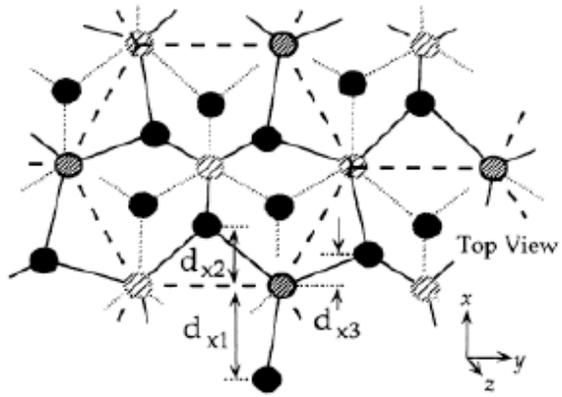
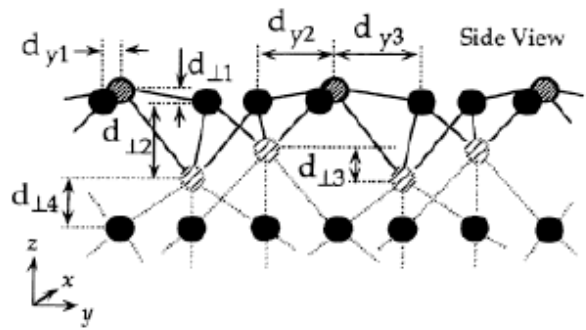

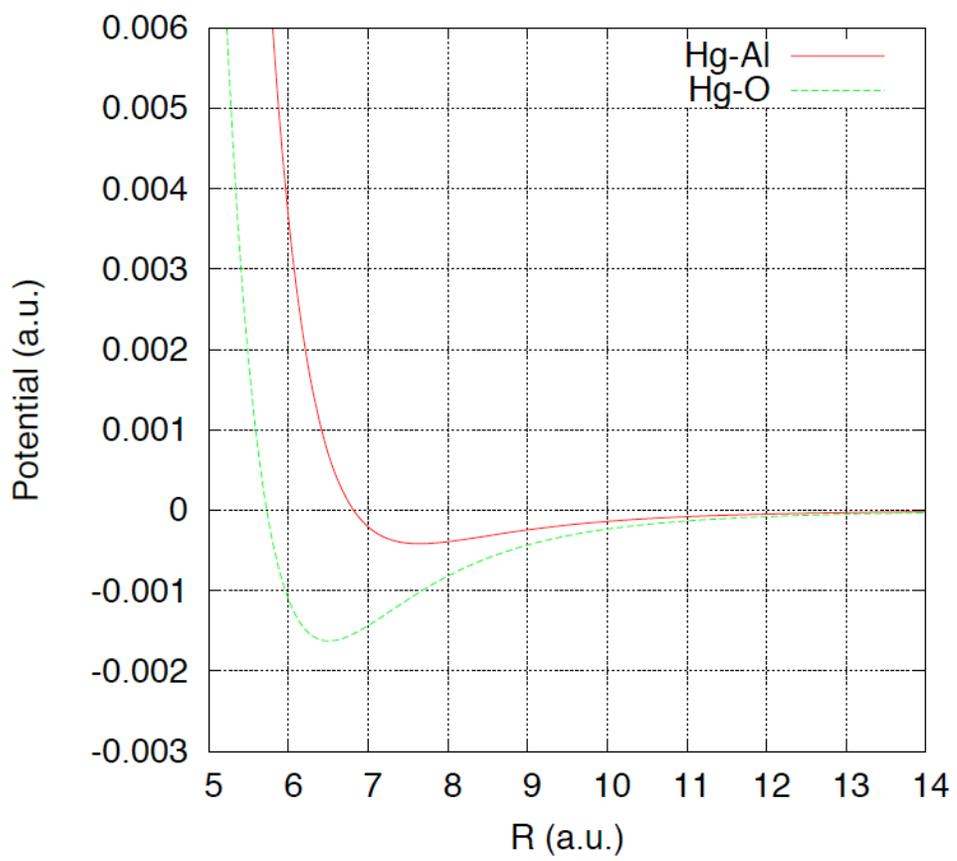

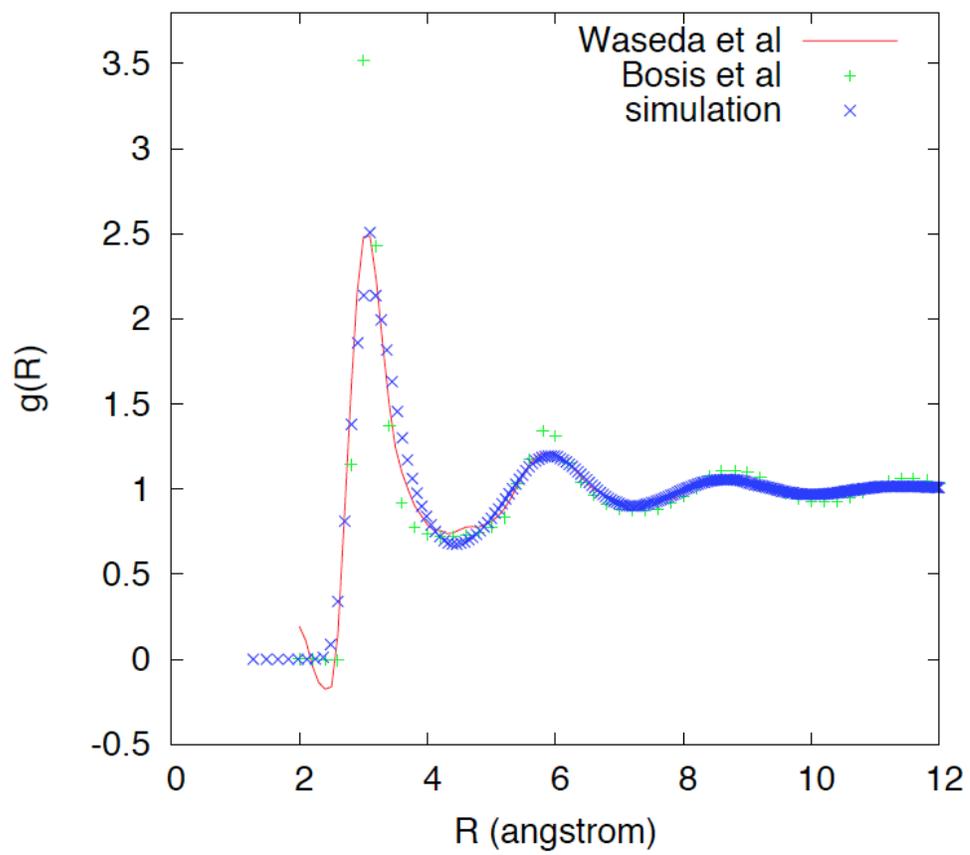

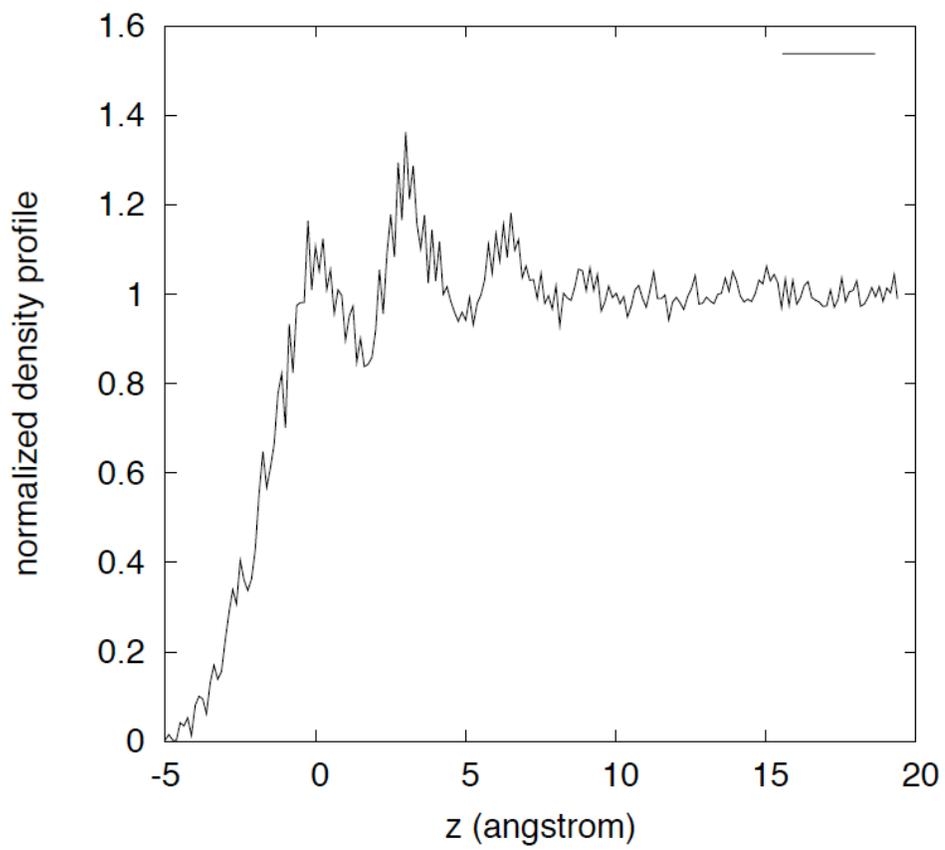

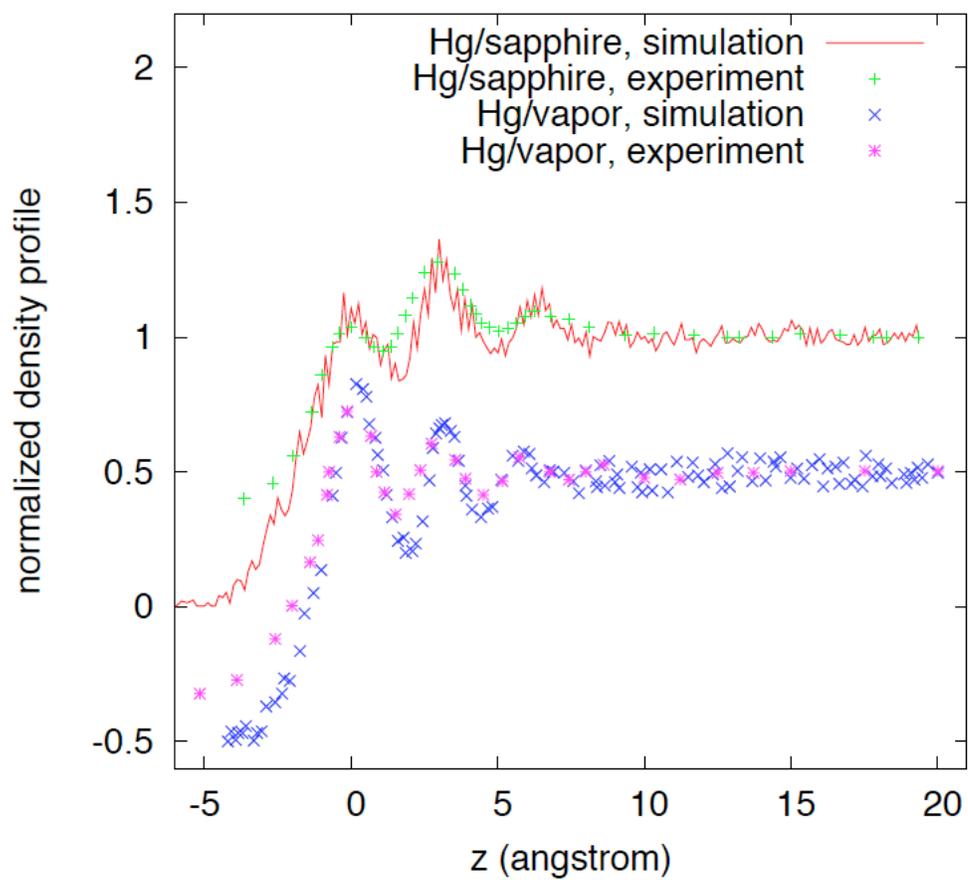